\documentclass[aps,prc,onecolumn,showpacs,superscriptaddress]{revtex4}
\usepackage[latin1]{inputenc}
\usepackage{bm}
\usepackage{epsfig}
\usepackage{amsthm}
\usepackage{amsfonts}
\usepackage{float}
\usepackage{amsmath,amssymb}
\usepackage{color}
\usepackage{slashed}
\usepackage{relsize}
\usepackage[toc,page]{appendix} 
\usepackage{wrapfig}
\usepackage{graphicx}
\usepackage{dcolumn}
\usepackage{bm}
\newcommand{\be}{\begin{equation}} 
\newcommand{\ee}{\end{equation}} 
\newcommand{\bea}{\begin{eqnarray}} 
\newcommand{\eea}{\end{eqnarray}} 
\newcommand{\nn}{\nonumber} 
\newcommand{\mintedim}[2]{{\int\kern-0.50em\mbox{{\small$\mathop{\frac{\mbox{{\small${\rm d^{#2}}\vect{#1}$}}}{\mbox{{\small$(2\pi)^{#2}$}}}}$}}\ }} 
\newcommand{\inteonedim}[1]{{\int_0^\infty\kern-1em\mbox{{\small${\rm d}{#1}$}}}} 
\newcommand{\vect}[1]{\bm{#1}} 

\begin{document}
\title{
\textcolor{black} {Nonextensive Boltzmann Transport Equation: the Relaxation Time Approximation and Beyond}
}

\author{Trambak Bhattacharyya}
\email{bhattacharyya@theor.jinr.ru}
\affiliation{Bogoliubov Laboratory of Theoretical Physics, Joint Institute for Nuclear Research, Dubna, 141980, \\
Moscow Region, Russia}

\begin{abstract} 
We derive approximate iterative analytical solutions of the nonextensive Boltzmann transport equation in the relaxation time approximation. 
The approximate solutions almost overlap with the exact solution for a considerably wide range of the parameter values found in describing
particle spectra originated in high-energy collisions. We also discuss the Landau kinetic approximation of the nonextensive Boltzmann transport 
equation and the emergence of the nonextensive Fokker-Planck equation, and use it to estimate the drag and diffusion coefficients of highly energetic 
light quarks passing through a gluonic plasma.
\end{abstract}

\maketitle
%

\section{Introduction}
Studying the transport properties of a medium is an important field of research which helps characterizing a system. 
One of the most widely used transport equations is the Boltzmann Transport Equation (BTE) \cite{degroot} derived long ago for a dilute gas 
of classical particles. Since then, it has been used in many fields including that of high energy collisions. For example, to understand 
the transport of particles inside the quark-gluon plasma medium produced in high energy collisions, either the BTE, or some approximated 
version of it has been used \cite{Yao:2020xzw,Qiao:2020yry,Kurian:2019nna,Singh:2018wps,Tripathy:2017nmo,tsallisraaepja,Mazumder:2011nj}. 
In this approach, the evolution of a distribution function is studied, and is utilized to understand the observables in the multi-particle 
production experiments colliding protons or heavy-ions. Investigating the properties of such a medium calls for a statistical approach which is conventionally based on the Boltzmann-Gibbs statistics  and the exponential single particle distribution which is obtained as a stationary state solution of the conventional BTE. It is, however, observed  that the BTE may be inadequate to describe freeze-out in high energy collision experiments \cite{Magas:2005mb}. Also, conditions leading to the 
Boltzmann-Gibbs statistics may not be satisfied because systems produced rather display long-range correlation and fluctuation. 
For strongly 
interacting plasmas like the QGP, the plasma parameter $\Gamma$ given by the ratio of the average potential to kinetic energy is larger than 1. In these situations, interaction length is much larger than the Debye screening length and a strongly interacting plasma like QGP does not follow the conventional kinetic equation leading to equilibration of system. It can be shown that for such a system, power-law stationary states like the one characterized by the Tsallis statistics arise \cite{rafelskiwaltonprl}.

Imprints of nonextensivity are also found in the hadronic transverse momentum distributions which are power-like, and hence, are far from being exponential. Such a power-like 
distribution which is routinely used in the field of high-energy collisions is proportional to the following factor \cite{jpg20,Cleymans:2013rfq},
\bea
\left(1+\frac{q-1}{T}E\right)^{-\frac{q}{q-1}},
\label{tsallisdist}
\eea
which originates from the Tsallis statistics \cite{Tsal88}. In Eq.~\eqref{tsallisdist}, $E=\sqrt{|\vec{P}|^2+m^2}$ is the single particle energy for a particle with three-momentum $\vec{P}$ and mass $m$. $q$ is the entropic parameter, and $T$ is the Tsallis temperature. The $q$ parameter is related to the relative variance in temperature (or number of particles) which is reminiscent of the fluctuating ambiance \cite{Wilk00,Wilk09}. Also, analyzing scaling properties of the Yang-Mills theory, $q$ is deduced from field theory parameters like $N_c$ (no. of colours) and $N_f$ (no. of flavours) \cite{deppmanprdq}. When, $q\rightarrow$1, the `Tsallis factor' (represented as the $q$-exponential $\exp_q$ raised to the power $q$) approaches being Boltzmann-like (exponential), {\it i.e.}
\bea
\lim_{q\rightarrow1} \left(1+\frac{q-1}{T}E\right)^{-\frac{q}{q-1}} = \lim_{q\rightarrow1} \left\{\exp_q \left(-\frac{E}{T}\right) \right\}^q= \exp\left(-\frac{E}{T}\right).
\label{tsallislim}
\eea
It is found that Eq.~\eqref{tsallisdist}, which describes the high-energy collision data, is the stationary solution of a modified Boltzmann transport equation 
\cite{lavagnopla,wilkosada,Biro:2012ix} inspired by the Tsallis statistics for a non-equilibrium distribution $f$, 
\bea
P^{\mu} \partial_{\mu} f^q= \mathcal{C}_q[f].
\label{NEBTE}
\eea
One of the modifications in the Tsallis-inspired Boltzmann transport equation in Eq.~\eqref{NEBTE}, henceforth to be called the nonextensive 
BTE (NEBTE), is that the distribution function at the left hand side is raised to the power $q$. This power is important for conservation laws to hold 
\cite{lavagnopla}. 
The collision term $\mathcal{C}_q$ is also modified to accommodate a generalization of the `molecular 
chaos' (stosszahlansatz) hypothesis which may not be valid in the system produced in high-energy collisions. It is to be noted that when $q\rightarrow1$, 
the NEBTE converges to the conventional BTE.

Analytical solutions of the conventional Boltzmann transport equation exist under certain approximations. One such example is the relaxation time approximation. Solution of the BTE under this approximation considering simplifying scenarios is well-studied and has been subject matters in recent studies involving the Tsallis-like distributions \cite{tsallisraaepja,maciej,Wilk:2021jpl}. In Ref.~\cite{Wilk:2021jpl}, authors also consider the time variation of the 
Tsallis $q$ parameter to go beyond the relaxation time approximation. 

Under the same simplifying assumptions of a homogeneous plasma with no external force, the NEBTE under the relaxation time approximation (RTA) is non-linear because of the power index $q$ in the l.h.s of Eq.~\eqref{NEBTE}. And hence, unlike the BTE, finding an exact, and closed analytical solution is difficult. In the present paper, we compute approximate iterative analytical solutions of the nonextensive BTE in the relaxation time approximation. We observe that the approximate solutions almost overlap with the exact solution for a considerably wide range of the parameter values. In addition to this, we also discuss
the Landau kinetic approximation of the NEBTE and the emergence of the nonextensive Fokker-Planck equation (see also Ref. \cite{hqtsfp,deppmannnefpe}). Later, this equation has been used to estimate the drag and diffusion coefficients of energetic light quarks passing through a gluonic plasma.


\section{nonextensive Boltzmann transport equation}
If we expand the l.h.s of Eq.~\eqref{NEBTE} assuming that there is no net external force, we obtain,
\bea
\frac{\partial f^q(\vec{P},\vec{r},t)}{\partial t} + \vec{v}.\vec{\nabla} f^q(\vec{P},\vec{r},t) = E^{-1}\mathcal{C}_q[f(\vec{P},\vec{r},t)],
\label{NEBTEexpand}
\eea
where for a particle with three-momentum magnitude $|\vec{P}|\equiv \sqrt{P_t^2+P_z^2}$ (where $P_t$ and $P_z$ are transverse and longitudinal momenta respectively), 
and energy $E$, velocity $v=|\vec{P}|/E$, and $\vec{r}$ denotes the position coordinates. Assuming an expansion along the collision axis which is taken to be the z-direction,
\bea
\frac{\partial f^q(\vec{P},z,t)}{\partial t} + v_z\frac{\partial f^q(\vec{P},z,t)}{\partial z}  = E^{-1}\mathcal{C}_q[f(\vec{P},z,t)].
\label{NEBTEz}
\eea
Under the Bjorken's assumption of boost-invariance at the central rapidity region, Eq.~\eqref{NEBTEz} can be written as \cite{Baym},
\bea
\left.\frac{\partial f^q(\vec{P}_t,P_z,t)}{\partial t} \right|_{P_zt}= E^{-1}\mathcal{C}_q[f(\vec{P}_t,P_z,t)].
\label{NEBTEBj}
\eea
For the binary collisions of particles with four-momenta $P\equiv(E,\vec{P})$ (having the momentum distribution $f$), and $Q\equiv(\mathcal{E},\vec{Q})$ (having the momentum distribution $g$), 
which become $P'\equiv(E',\vec{P'})$, and $Q'\equiv(\mathcal{E'},\vec{Q'})$ after interaction, the collision term of the nonextensive Boltzmann transport equation can be written as \cite{wilkosada,Biro:2012ix},
\bea
\mathcal{C}_q
&=& \frac{1}{2} \int \frac{d^3\vec{Q}}{(2\pi)^3\mathcal{E}} \frac{d^3\vec{Q'}} {(2\pi)^3\mathcal{E'}} \frac{d^3\vec{P'}} {(2\pi)^3 E'}  |\overline{M}|^2 (2\pi)^4 
\delta^4(P+Q-P^{'}-Q^{'}) 
\left[ h_q \{f(\vec{P'}),g(\vec{Q'})\} -h_q \{f(\vec{P}),g(\vec{Q})\} \right].
\label{necollterm} \nn\\
\eea
$M$ is the amplitude of the quark-quark or quark-gluon collisional processes in a quark-gluon plasma medium.
The function $h_q$ is a generalization of the molecular chaos hypothesis, and combines the two distributions $f$, and $g$ in the following 
way,
\bea
h_q\{f,g\} = \exp_q \left[\log_q (f)+\log_q(g)\right],
\label{hqdef}
\eea
where
\bea
\log_q f = \frac{1-f^{1-q}}{q-1}.
\eea
Here $\log_q$ is the $q$-logarithm function and becomes the conventional logarithm in the limit $q\rightarrow1$,
which also implies, 
\bea
\lim_{q\rightarrow1}h_q\{f,g\} &=& \lim_{q\rightarrow1} \exp_q \left[\log_q (f)+\log_q(g)\right] \nn\\
&=& \exp \left[\log (f)+\log (g)\right] \nn\\
&=& f\times g,
\label{hqdef}
\eea
{\it i.e.}, the distribution functions are independent, which is a consequence of the molecular chaos hypothesis. In this limit, one gets back the
collision term of the conventional BTE \cite{svetitsky},
\bea
\mathcal{C}
&=& \frac{1}{2} \int \frac{d^3\vec{Q}}{(2\pi)^3\mathcal{E}} \frac{d^3\vec{Q'}} {(2\pi)^3\mathcal{E'}} \frac{d^3\vec{P'}} {(2\pi)^3 E'}  |\overline{M}|^2 (2\pi)^4 
\delta^4(P+Q-P^{'}-Q^{'}) 
\left[ f(\vec{P'}) g(\vec{Q'}) - f(\vec{P})g(\vec{Q}) \right].
\label{necollterm} \nn\\
\eea

The form of the collision term in Eq.~\eqref{necollterm} is motivated by 
the fact that in strongly interacting systems, one needs a transport equation whose stationary solution is given by a 
power-law distribution. The conventional Boltzmann transport equation that considers stosszahlansatz in the collision term yields only an exponential
distribution and a generalization leads to a Tsallis power-law stationary solution. The generalized stosszahlansatz indicates an interplay between
the probe particle and medium particle distributions. In this article, this interplay has been characterized using the $q$ functions. 
A deformed Fokker-Planck equation (that will be considered in a later section), just like the generalized Boltzmann transport equation, also leads to a power-law
stationary solution and can adequately describe traversal of highly energetic particles inside strongly interacting QCD medium. 

\section{nonextensive Boltzmann transport equation in the relaxation time approximation}
If at time $t=0$, all the external forces are switched off and the gradient is cancelled, the nonextensive Boltzmann 
transport equation for the distribution $f$ in the relaxation time approximation is given by \cite{silvanebterta} (except for 
the power $q$, as discussed in \cite{lavagnopla}), 

\bea
\frac{\partial f^q}{\partial t} &=& - \frac{\left(f-f_{\text{eq}}\right)} {\tau} \nn\\
\frac{\partial f}{\partial t}    &=& - \frac{ \left(f^{2-q} - f_{\text{eq}} f^{1-q}\right)}   {q\tau},
\label{nebterta}
\eea
where $\tau$ is the relaxation time. Integrating the above equation,
\bea
\int \frac{df}{\left(f^{2-q} - f_{\text{eq}} f^{1-q}\right)} &=&  \mathcal{K}-\theta  \nn\\
\frac{1}{q-1}\int \frac{dw} { \left(1- f_{\text{eq}} w^{-\frac{1}{q-1}} \right) } &=&  \mathcal{K}-\theta,\quad \nn\\
\text{where} \quad w\equiv f^{q-1}, \quad \theta = \frac{t}{q\tau},
\eea
and $\mathcal{K}$ is the integration constant which may be obtained from the boundary condition, $f(t=0)=f_{\text{in}}$, where $f_{\text{in}}$ is the
initial distribution. We expand the integrand in a negative binomial series and integrate.

\bea
\frac{1}{q-1} \int dw \left( 1 + f_{\text{eq}} w^{-\frac{1}{q-1}} + f_{\text{eq}}^2 w^{-\frac{2}{q-1}} +...\right) &=&  \mathcal{K}-\theta 
\quad \left( \left|f_{\text{eq}} w^{-\frac{1}{q-1}}\right| \equiv \left|\frac{f_{\text{eq}}}{f}\right| < 1 \right)
\nn\\ 
\Rightarrow \frac{f^{q-1}}{q-1} \mathlarger{\mathlarger{\sum}}_{s=0}^{\infty} \frac{(1)_s(1-q)_s}{s!(2-q)_s} \left(\frac{f_{\text{eq}}}{f}\right)^s
&=&  \mathcal{K} -\theta  \nn\\
\Rightarrow \frac{f^{q-1}}{q-1}   \, _2F_1\left(1,1-q;2-q;\frac{f_{\text{eq}}}{f}\right) &=&  \mathcal{K}-\theta, \nn\\
\label{nebtesol}
\eea
where $`(.)_s$' in the second line is the rising Pochhamer symbol given by \cite{abst},
\bea
(a)_s = 
\begin{cases}
1 & s=0 \\
a(a+1).....(a+s-1) & \forall s>0,
\end{cases}
\eea
and $\, _2F_1$ is the hypergeometric function \cite{Bateman}. The integration constant is given by,
\bea
\mathcal{K} = \frac{f_{\text{in}}^{q-1}}{q-1}   \, _2F_1\left(1,1-q;2-q;\frac{f_{\text{eq}}}{f_{\text{in}}}\right).
\eea
Hence, the solution of the nonextensive Boltzmann transport equation in the relaxation time approximation may be obtained once we solve
Eq.~\eqref{nebtesol} for $f$. Although the solution of Eq.~\eqref{nebtesol} can be found using numerical methods, in this paper we calculate 
approximate analytical expressions for the solutions using the series expansion of the hypergeometric function given in the second line of 
Eq.~\eqref{nebtesol}. 

It is straightforward to find the zeroth order solution of Eq.~\eqref{nebtesol} ({\it i.e.} for $s=0$) from the following equation,
\bea
\Psi_0 &=& f^{q-1}-(q-1)\left( \mathcal{K} -\theta \right)=0\nn\\
\Rightarrow f_0 &=& \left[(q-1)\left( \mathcal{K} -\theta \right)\right]^{\frac{1}{q-1}}.
\label{sol0}
\eea
The first order equation, whose solution we denote by $f_1(t)$, is given by,
\bea
\Psi_1 &=& f^{q-1}+ \left(\frac{1-q}{2-q} \right) f_{\text{eq}} f^{q-2} - (q-1)\left( \mathcal{K} -\theta \right) = 0.
\label{sol1}
\eea
And the exact solution, which we denote by $f_{\text{e}}(t)$, is obtained from the following equation,
\bea
\Psi_{\text{e}} = \frac{f^{q-1}}{q-1}   \, _2F_1\left(1,1-q;2-q;\frac{f_{\text{eq}}}{f}\right) -  \mathcal{K}-\theta =0
\label{solex}
\eea
Though it is straightforward to obtain the zeroth order solution using the inverse function, for the first order and beyond, it becomes difficult.

\begin{figure*}[!htb]
\vspace*{+0cm}
\minipage{0.43\textwidth}
\includegraphics[width=\linewidth]{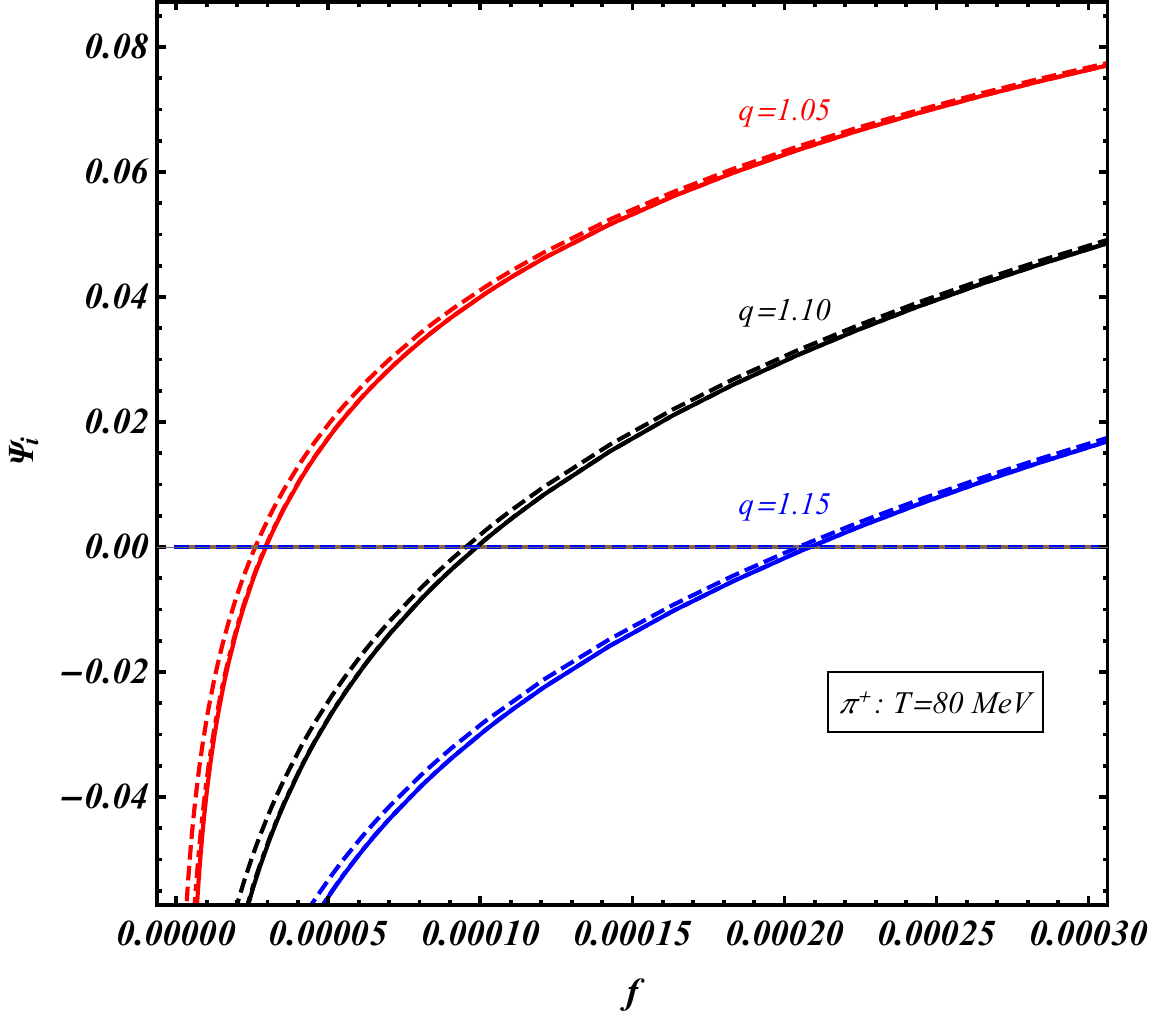}
\endminipage\hfill
\minipage{0.43\textwidth}
\vspace*{+0cm}
\hspace*{-3cm}
\includegraphics[width=\linewidth]{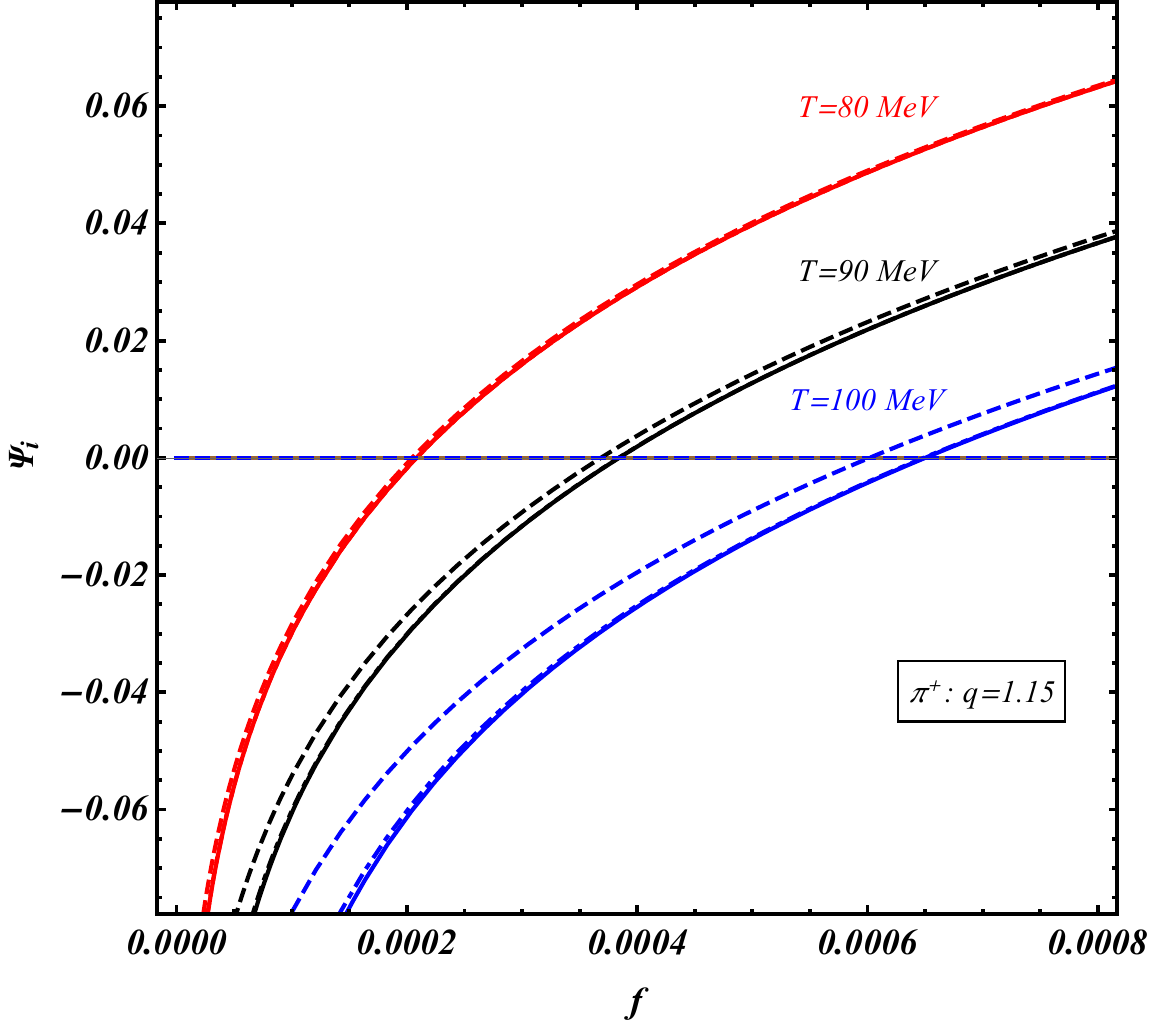}
\endminipage\hfill
\caption{Graphical solutions of Eqs.~\eqref{sol0}-\eqref{solex} for different $q$ and $T$ values, mass $m=139.57$ MeV (pion mass), $p_{\text{T}}$=1 GeV, $q\theta=0.11$.}
\label{fig1}
\end{figure*}

In what follows, we propose an approximate analytical first order solution for the nonextensive Boltzmann transport equation
in the relaxation time approximation. This proposal is based on Fig.~\ref{fig1} which finds the graphical solutions of 
Eqs.~\eqref{sol0},~\eqref{sol1}, and~\eqref{solex}, which are located at the points where $\Psi_i^s$ ($i=0,1$, e) change sign. 
We observe that at the transverse momentum values near 1 GeV and above, the solution of the zeroth order equation (dashed line) is very close to that of the 
exact equation (solid line) which entirely overlaps with the solution of the first order equation (dot-dashed line, not always visible because of overlapping). 
Hence, we propose to write the solution of the first order equation as a tiny increment over that of the zeroth order in the following way,
\bea
f_1=f_0+\epsilon_1, \quad |\epsilon_1| <<f_0.
\label{solf1eps}
\eea
Afterwards, we put Eq.~\eqref{solf1eps} in Eq.~\eqref{sol1}, expand in terms of $\epsilon_1$ up to the first order (since $\epsilon_1$ is a small quantity), solve for $\epsilon_1$ 
and get $f_1$ in terms of $f_0$ whose analytical form is already known from Eq.~\eqref{sol0}. This gives us the following expression for the solution of the
first order equation,
\bea
f_1 \approx f_0 + \frac{f_0}{f_0+f_{\text{eq}}} \left[ \frac{f_{\text{eq}}}{2-q} + \frac{f_0}{1-q} + f_0^{2-q} \left( \mathcal{K} -\theta \right) \right]. \nn\\
\label{nebtertaapproxsol}
\eea
\begin{wrapfigure}{l}{0.40\textwidth}
  \vspace{0pt}
   \centering
   {\includegraphics[width=0.40\textwidth]{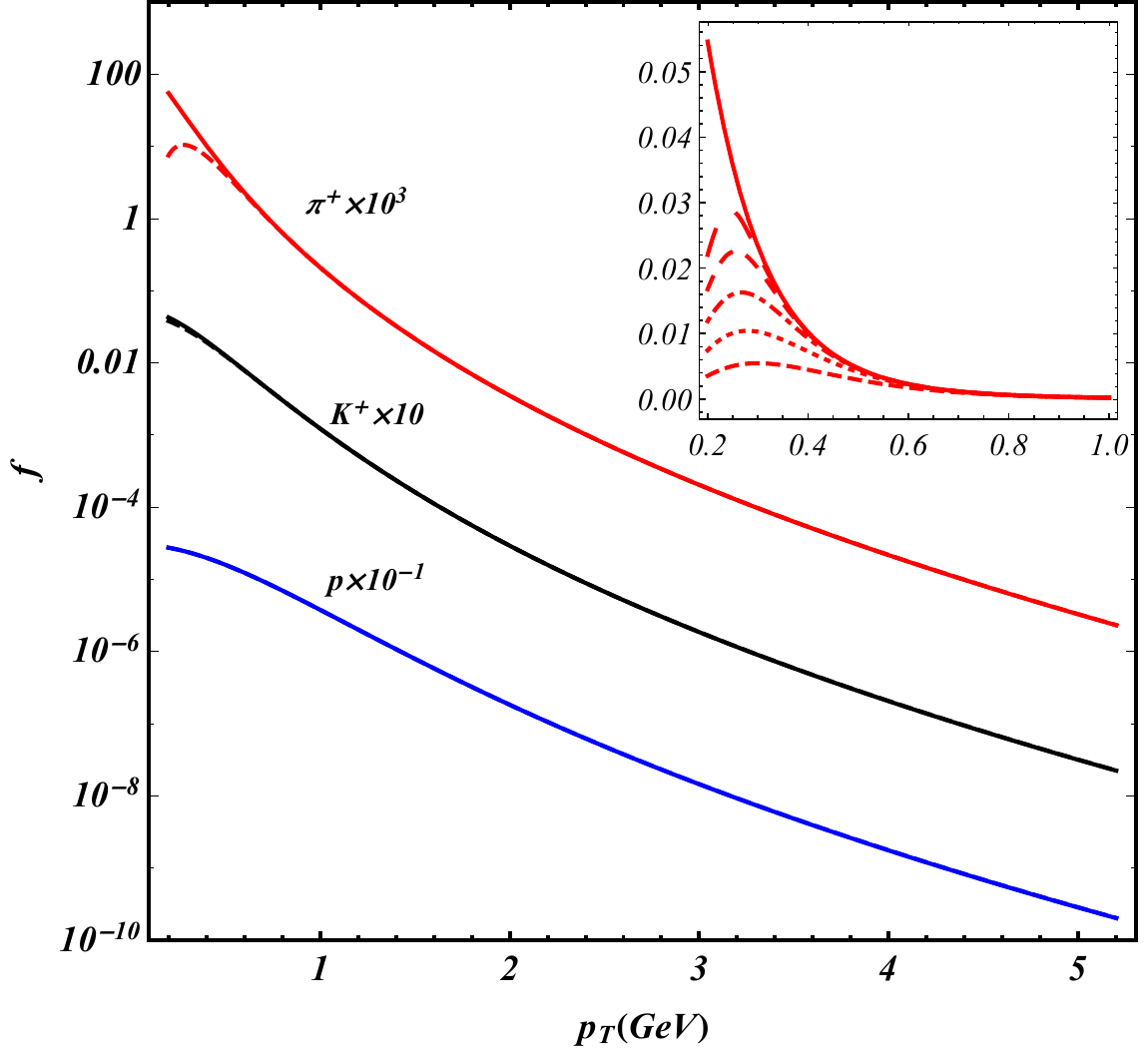}}
  \vspace{-20pt}
\caption{Comparison between the exact and the first order approximated solutions. $q=1.15,~T=80$ MeV, and $\theta=0.096$.}
\label{fig2}
 \vspace{-5pt}
\end{wrapfigure}
Next, we find out how good the approximate solution Eq.~\eqref{nebtertaapproxsol} is, and if there are any limitations of the approximation. To do this,
we compare the exact numerical solution of Eq.~\eqref{nebterta} and the approximate solution given by Eq.~\eqref{nebtertaapproxsol}
in Fig.~\ref{fig2} for $q=1.15,~T=80$ MeV, and $\theta=0.096$. We use the mass values of the pions, kaons and protons.  We find that for the pions 
having very low transverse momentum ($\lesssim$ 0.6 GeV), the approximated solution Eq.~\eqref{nebtertaapproxsol} underestimates the exact 
solution. For mass values lower than the pion mass, this disagreement at very low transverse momentum remains. However, for the most part of the 
pion spectrum, the first order approximation works really well. For heavier particles like the kaons and the protons, the first order approximated solution 
is as good as the exact solution for the whole spectrum.
In the inset of Fig.~\ref{fig2} we plot the zeroth order (bottommost) to the fourth order (second from the top)  solutions (for a particle with pion mass) along with the exact solution (red, solid). We observe that higher order solutions lead to a greater degree of agreement between the approximate analytical solution and the exact solution at very low momentum. Following Eq.~\eqref{solf1eps}, the first order and the higher order solutions can be represented as,
\bea
~~~~~~~~~~~~~~~~~~~~~~~~~
f_i = f_{i-1} +\epsilon_i, \quad i=1,~2,~3,...,
\eea
where $\epsilon_i^{\text{s}}$ are calculated from the following equation,
\bea
\epsilon_i = \frac{ f_{i-1}  }
{  \mathlarger{\sum}_{r=0}^i  f_{\text{eq}}^r f_{i-1}^{i-r} }
\left( f_{i-1}^{i+1-q} (\mathcal{K}-\theta) + \mathlarger{\mathlarger{\sum}}_{r=0}^i \frac{f_{\text{eq}}^r f_{i-1}^{i-r}} {r+1-q} \right). \nn\\
\eea


\vspace{-30pt}

\section{Nonextensive Fokker-Planck equation and its relation to the nonextensive Boltzmann transport equation}
The Stratonovich form of the nonextensive Fokker-Planck equation for a function $f'({\bold x})$, where ${\bold x} \in \mathbb{R}^N$ 
is a dimensionless state variable, is given by \cite{borland},
\bea
\frac{\partial f'}{\partial t} = - \frac{\partial}{ \partial x} \left(K(x) f'\right) + \frac{\partial}{ \partial x} \left(D(x) \frac{\partial f'^{2-q}}{ \partial x} \right)
\label{nefpex}
\eea
in one dimension. $K$, and $D$ are drift and diffusion. 
Putting $\partial f' / \partial t =0 $, the normalized stationary solution of Eq.~\eqref{nefpex} is,
\bea
f' = \Gamma \left[1-\beta (1-q) V_{\text{eff}} (x)\right]^{\frac{1}{1-q}},
\eea
where $\beta$, and $\Gamma$ are appropriate constants with
\bea
V_{\text{eff}} (x) = - \int \frac{K(x)}{D(x)} dx.
\eea
For the drift and diffusion terms,
\bea
K(x) = -c_K x^{\gamma}; \quad D(x) = c_D x^r,\quad \text{and} \quad s=\gamma-r+1,
\eea
The ansatz for a time ($t$)-dependent solution is written to be \cite{borland}
\bea
f' (t)= \Gamma(t) \left[1-\beta(t) (1-q) x^s\right]^{\frac{1}{1-q}}.
\label{borlandansatz}
\eea
By putting the ansatz in Eq.~\eqref{nefpex}, the time dependence of $\beta$, and $\Gamma$ can be calculated.

It is, however, also possible to get the nonextensive Fokker-Planck equation from Eq.~\eqref{NEBTEBj} when
the Landau approximation is imposed \cite{landau}. The Landau approximation is motivated by the fact that the most of the
parton-parton collisions are soft. That means that the collision rate is the highest near the three-momentum transfer 
$\vec{k}=\vec{P}-\vec{P'}=\vec{Q'}-\vec{Q}\approx0$. To establish this equation, we consider the passage of high-energy light quarks, 
whose distributions evolve like Eq.~\eqref{NEBTEBj}, through a gluonic medium which is thermalized in the Tsallis sense at a 
temperature $T_g$. The gluonic distribution $g(Q)$ is given by
\bea
g = \left(1+(q-1)\frac{|\vec{Q}|}{T_g}\right)^{-\frac{1}{q-1}}.
\label{gldist}
\eea
We also assume that the shape of the evolving hard quark distribution is dictated by the Tsallis-like function so that,
\bea
f = \left(1+(q-1)\frac{E}{T}\right)^{-\frac{1}{q-1}}.
\label{qdist}
\eea
Using Eqs.~\eqref{gldist}, and \eqref{qdist} in the definition Eq.~\eqref{hqdef}, the collision term reads,
\bea
\mathcal{C}_q
&=& \frac{1}{2} \int \frac{d^3\vec{Q}}{(2\pi)^3 \mathcal{E}} \frac{d^3\vec{Q'}} {(2\pi)^3 \mathcal{E}'} \frac{d^3\vec{P'}} {(2\pi)^3 E'}  |\overline{M}|^2 (2\pi)^4 
\delta^4(P+Q-P^{'}-Q^{'}) \times 
\nn\\
&& 
\left[ \left(1+(q-1)\frac{E'}{T}+ (q-1)\frac{|\vec{Q'}|}{T_g} \right)^{-\frac{1}{q-1}} -
\left(1+(q-1)\frac{E}{T}+ (q-1)\frac{|\vec{Q}|}{T_g} \right)^{-\frac{1}{q-1}} \right]
\nn\\
&=& \frac{1}{2} \int \frac{d^3\vec{Q}}{(2\pi)^3\mathcal{E}} \frac{d^3\vec{Q'}} {(2\pi)^3\mathcal{E'}} \frac{d^3\vec{P'}} {(2\pi)^3 E'}  |\overline{M}|^2 (2\pi)^4 
\delta^4(P+Q-P^{'}-Q^{'})  \left[f_{|\vec{P}-\vec{k}|} \mathcal{S}_{|\vec{P}-\vec{k}|,|\vec{Q}+\vec{k}|} 
- f_{|\vec{P}|} \mathcal{S}_{|\vec{P}|,|\vec{Q}|}
\right],
\label{necollterm1} \nn\\
\eea
where $f_{\vec{P}} \equiv f$, $E'=\sqrt{|\vec{P'}|^2+m^2}=\sqrt{|\vec{P}-\vec{k}|^2+m^2}$, and $f_{|\vec{P}-\vec{k}|}$ is the corresponding distribution. 
$\vec{Q'}=\vec{Q}+k$, and
\bea
\mathcal{S}_{|\vec{P}|,|\vec{Q}|} = 
\left(
1+
\frac{ (q-1)\frac{|\vec{Q}|} {T_g}  }
{1+(q-1)\frac{E}{T} }
\right)          
^{-\frac{1}{q-1} }.
\eea
Expanding Eq.~\eqref{necollterm1} around $\vec{k} \approx 0$, we obtain the Ito form \cite{borland} of the nonextensive Fokker-Planck equation from Eq.~\eqref{NEBTEBj},
\bea
\frac{\partial f}{\partial t} = - \frac{\partial}{ \partial P_i} \left(A_{i,q} f\right) + \frac{\partial}{ \partial P_i \partial P_j} \left(B_{ij,q} f^{2-q} \right)
~~~~~~~~(i=1,2,3),
\label{nefpeIto}
\eea
where $A_{i,q}$, and $B_{ij,q}$, the nonextensive Fokker-Planck drag and diffusion coefficients are given by,
\bea
A_{i,q} &=& \frac{1}{2Eq} \int \frac{d^3\vec{Q}}{(2\pi)^3\mathcal{E}} \frac{d^3\vec{Q'}} {(2\pi)^3\mathcal{E'}} \frac{d^3\vec{P'}} {(2\pi)^3 E'} |\overline{M}|^2 (2\pi)^4 
\delta^4(P+Q-P^{'}-Q^{'}) f^{1-q}S_{\vec{P},\vec{Q}} (P-P')_i \equiv \left<\left<(P-P')_i\right>\right>_q \nn\\
B_{ij,q} &=& \frac{f^{q-1}}{2} \left<\left<(P-P')_i (P-P')_j\right>\right>_q.
\label{tslightdrdiff}
\eea
For a similar approach following the Boltzmann-Gibbs statistics see \cite{jalqprd}.

\vspace{-0pt}
\section{Results}
\subsection{NEBTE in the RTA}
\begin{figure}[!htb]
\minipage{0.40\textwidth}
\includegraphics[width=\linewidth]{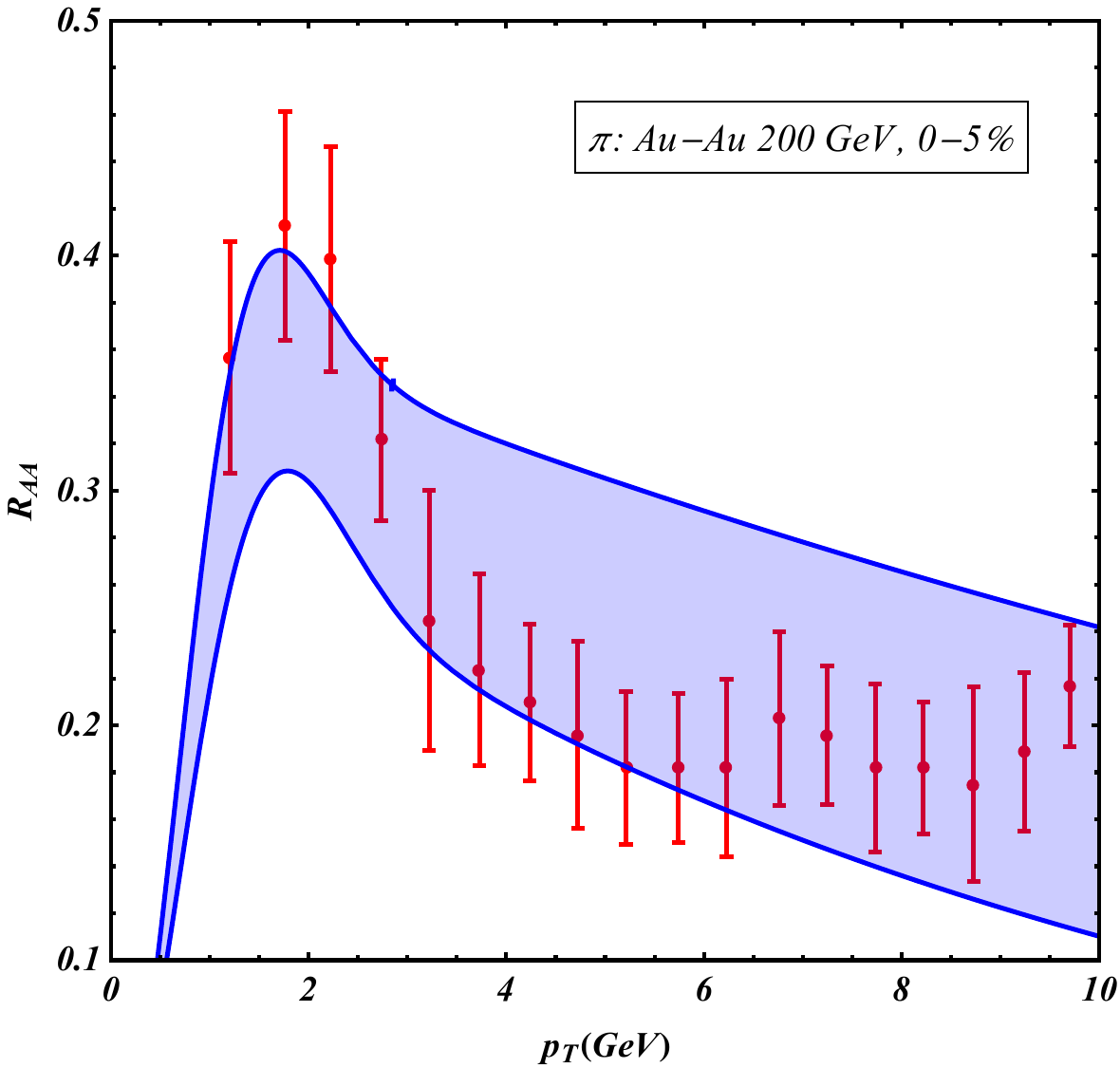}
\caption{$R_{\text{AA}}$ of neutral pions. $q=1.005\pm 0.003$, $T=0.066\pm0.024$ GeV, $t/\tau = 1.047\pm 0.096$, $\chi^2/\text{NDF}=18.64/15$.}\label{polar1}
\endminipage\hfill
\minipage{0.41\textwidth}
\includegraphics[width=\linewidth]{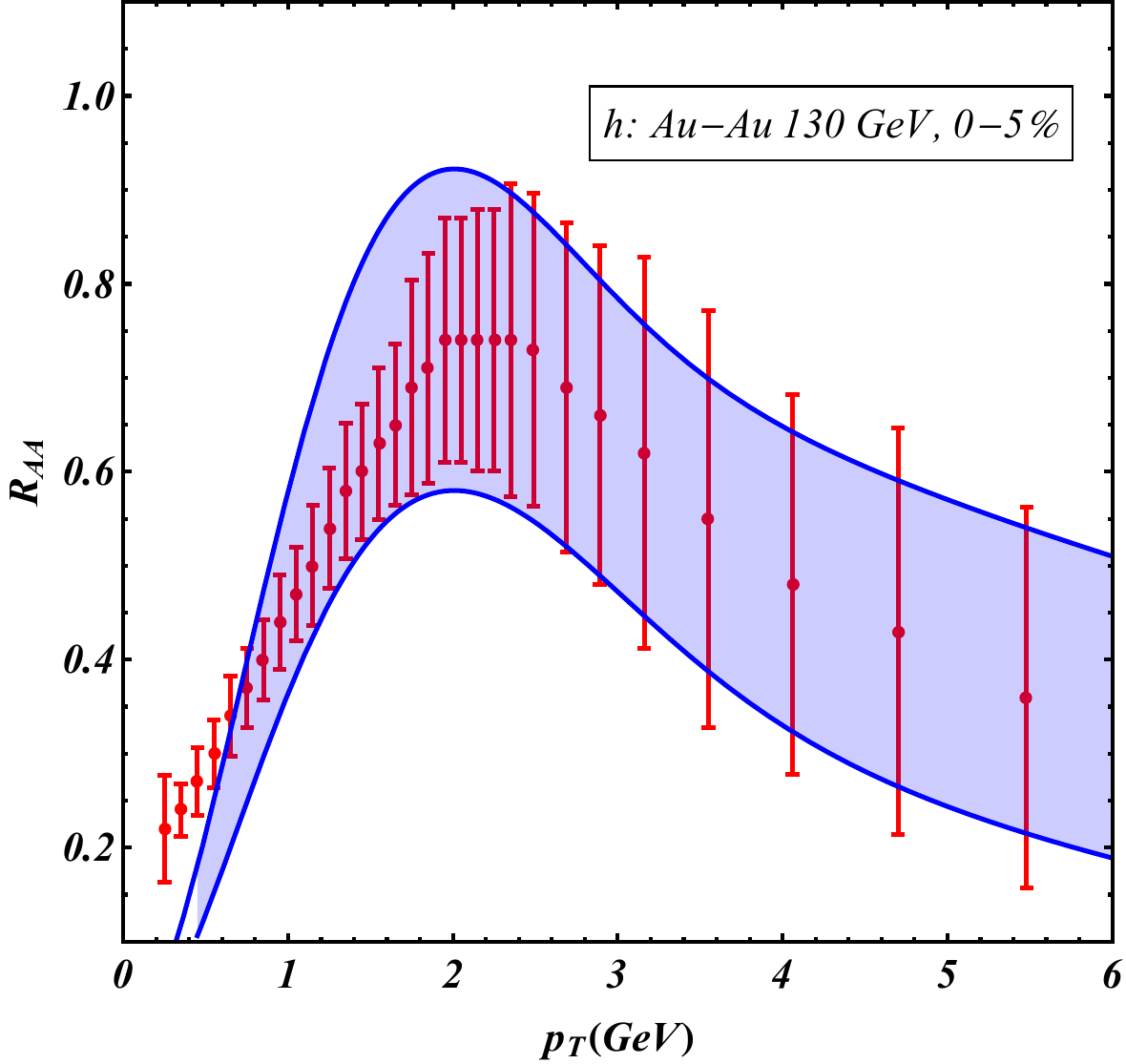}
\caption{$R_{\text{AA}}$ of all charged hadrons. $q=1.024\pm 0.011$, $T=0.187\pm0.050$ GeV, $t/\tau = 1.183\pm 0.0104$, $\chi^2/\text{NDF}=19.09/27$.}\label{polar3}
\endminipage\hfill
\end{figure}
We can use the solution $f_1$  given by Eq.~\eqref{nebtertaapproxsol} to describe (for a similar approach considering the conventional BTE see 
\begin{wrapfigure}{l}{0.40\textwidth}
  \vspace{-2pt}
   \centering
   {\includegraphics[width=0.40\textwidth]{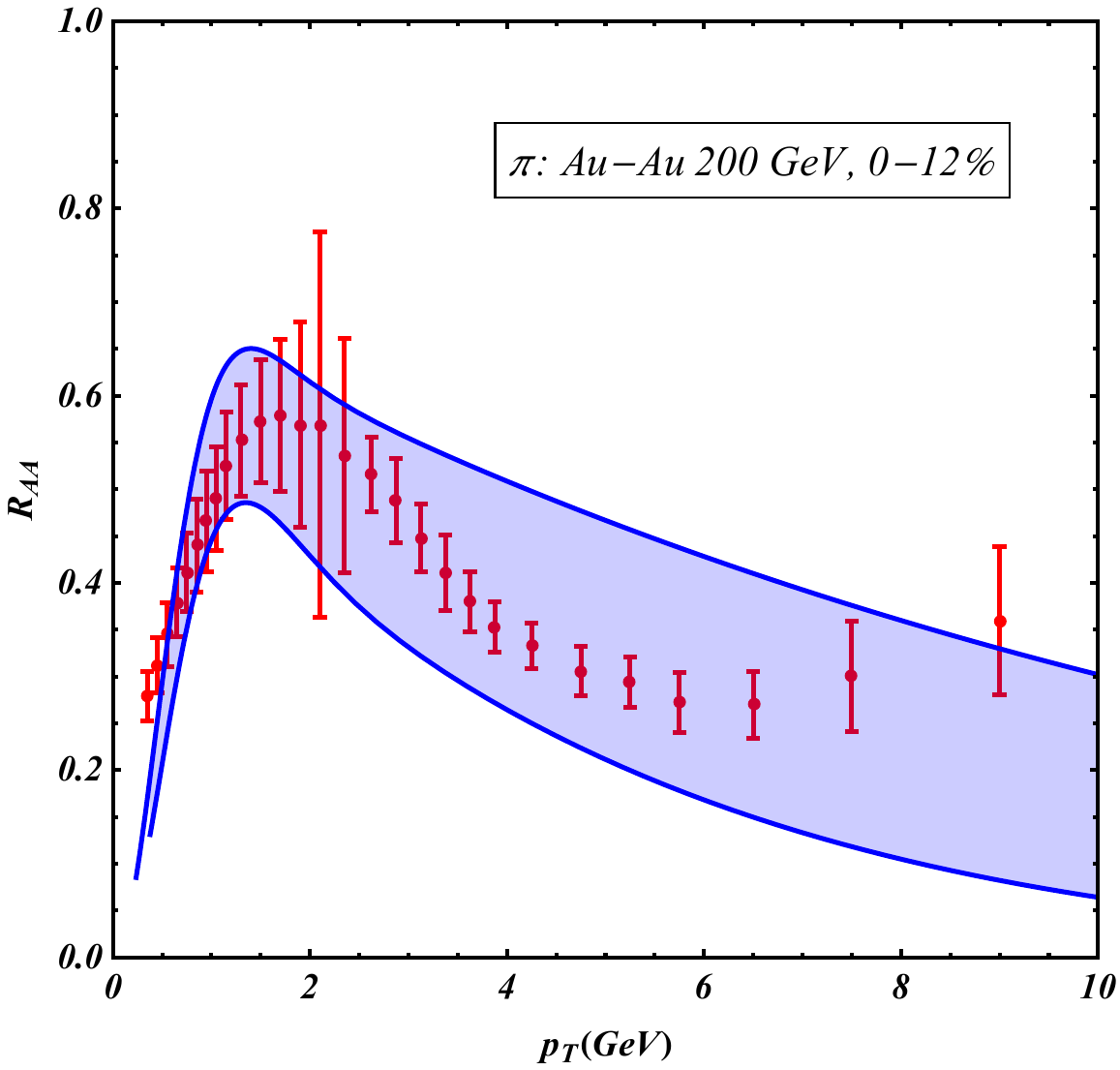}}
  \vspace{-25pt}
\caption{$R_{\text{AA}}$ of charged pions. $q=1.049\pm 0.020$, $T=0.156\pm0.036$ GeV, $t/\tau = 0.394\pm 0.065$, $\chi^2/\text{NDF}=23.38/25$.}\label{chpiraa}
\vspace{-45pt}
\end{wrapfigure}
Ref. \cite{tsallisraaepja}) an experimental observable like the nuclear suppression factor $R_{\mathrm{AA}}$ which is experimentally defined by,
\be
R_{\mathrm{AA}} = \frac{\left(d^2N/dp_{\mathrm{T}}dy\right)^{\mathrm{A+A}}}
{N_{\mathrm{coll}}\times\left(d^2N/dp_{\mathrm{T}}dy\right)^{\mathrm{p+p}}}.
\ee
$N_{\mathrm{coll}}$ is the number of nucleon-nucleon binary collisions (p+p) while a nucleus `A' collides with another nucleus. $d^2N/dp_{\mathrm{T}}dy$
denotes the differential yield within the transverse momentum and rapidity range $p_{\rm{T}}$ to $p_{\mathrm{T}}+dp_{\mathrm{T}}$ and $y$ to $y+dy$. 
If the yield from the nucleus-nucleus collisions would have been a linear superposition of nucleon-nucleon collisions, $R_{\mathrm{AA}}$ would have been 1. Deviation
of $R_{\mathrm{AA}}$ from 1 signifies medium modification. Theoretically, $R_{\mathrm{AA}}$ of the hadrons (`h') can be computed from the following equation,
\be
R_{\mathrm{AA}} = \frac{\sum_{\text{a}} \int f_{1}^\text{a}(p_{\text{a}},z)\left.\right|_{p_{\text{a}}=p_{\text{T}}/z} D_{\text{a}/\text{h}}(z) dz
}{\sum_{\text{a}} \int f_{\text{in}}^{\text{a}}(p_{\text{a}},z)\left.\right|_{p_{\text{a}}=p_{\text{T}}/z} D_{\text{a}/\text{h}}(z) dz},
\label{raatheo}
\ee
where $p_{\text{a}}$ is the momentum of the partons (denoted by `a') which fragment to the hadrons carrying the $z$ fraction of the partonic 
momentum. $D_{\text{a}/\text{h}}$ is the fragmentation function \cite{fragnpb,fragprd}. Description of $R_{\mathrm{AA}}$ data with the help of Eq.~\eqref{raatheo} is given in Figs.~\ref{polar1} (for the neutral pions \cite{pi0phenix200}), \ref{polar3} (for all the charged hadrons \cite{starhad130}), and \ref{chpiraa} (for the charged pions \cite{chpistar200}). The shaded regions in the figures correspond to the error bars in the parameter values obtained in the fitting.

\subsection{Nonextensive Fokker-Planck transport coefficients of the light quarks}
\begin{figure}[!htb]
\minipage{0.42\textwidth}
\includegraphics[width=\linewidth]{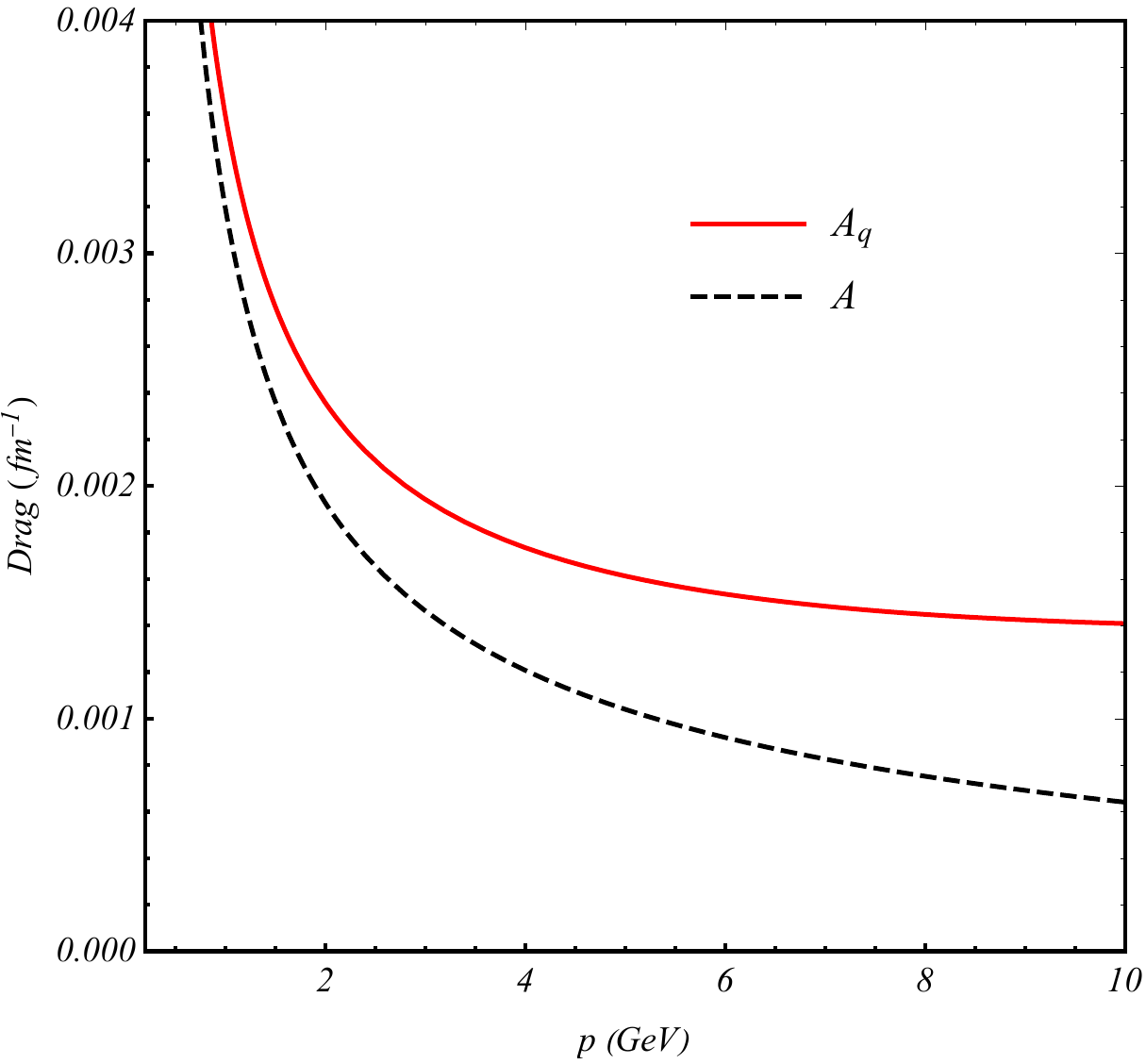}
\caption{Momentum variation of the nonextensive and extensive drag coefficients of the light quarks.}\label{dragp}
\endminipage\hfill
\minipage{0.42\textwidth}
\hspace*{-1cm}
\includegraphics[width=\linewidth]{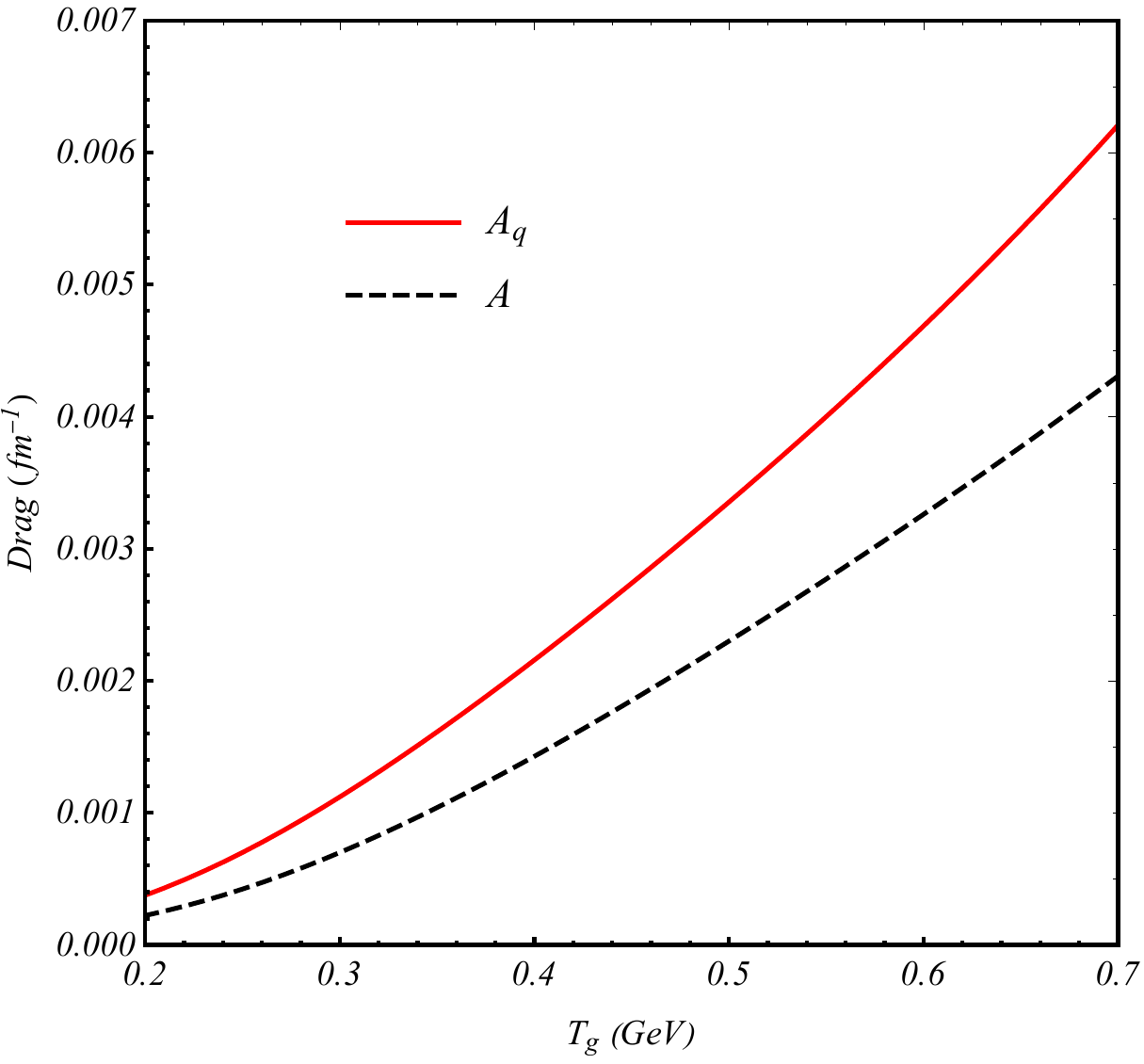}
\caption{Temperature variation of the nonextensive and extensive drag coefficients of the light quarks.}\label{dragT}
\endminipage\hfill
\minipage{0.42\textwidth}
\hspace*{-0cm}
\includegraphics[width=\linewidth]{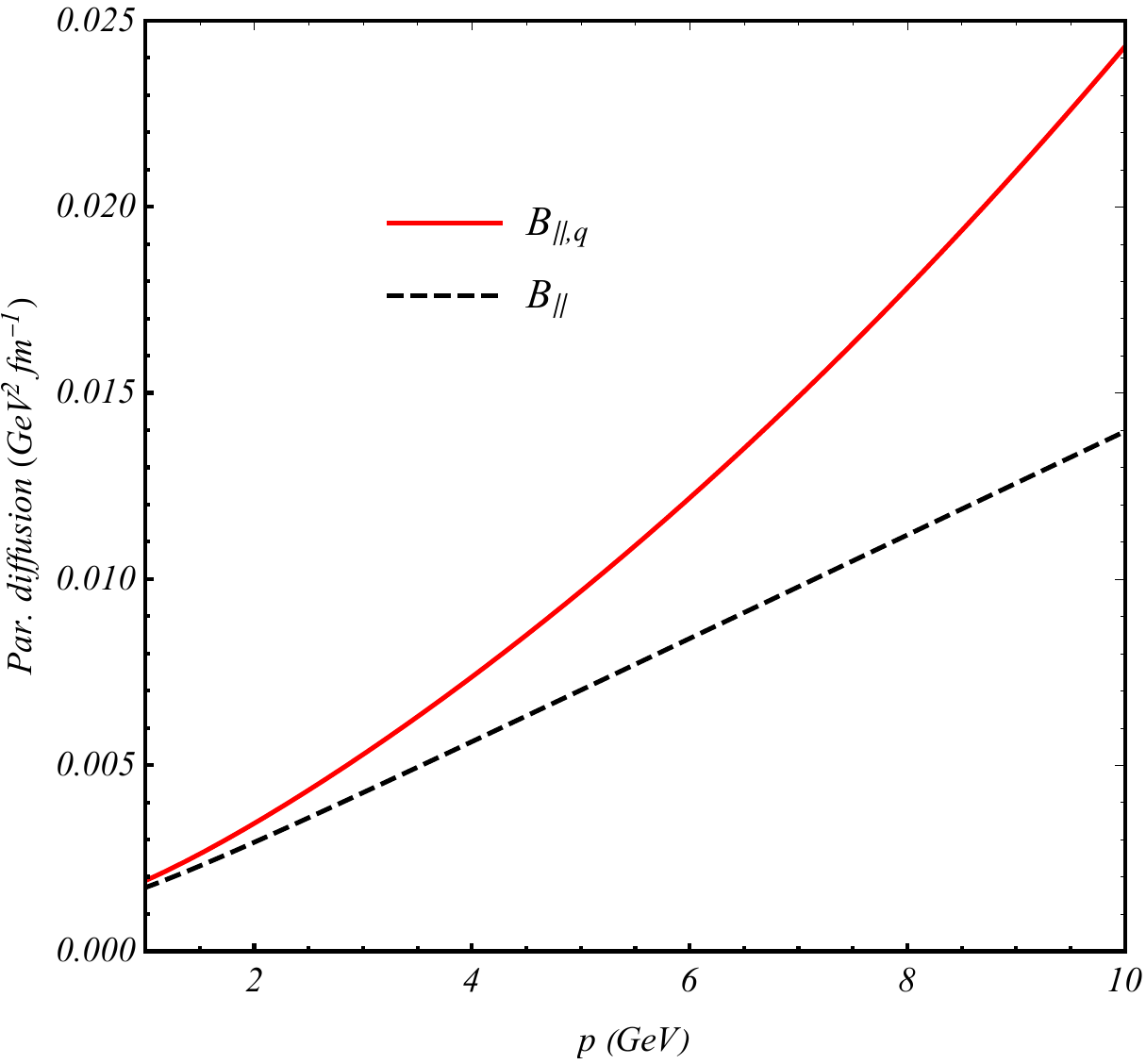}
\caption{Momentum variation of the nonextensive and extensive parallel diffusion coefficients of the light quarks.}\label{diffparp}
\endminipage\hfill
\minipage{0.42\textwidth}
\hspace*{-1cm}
\includegraphics[width=\linewidth]{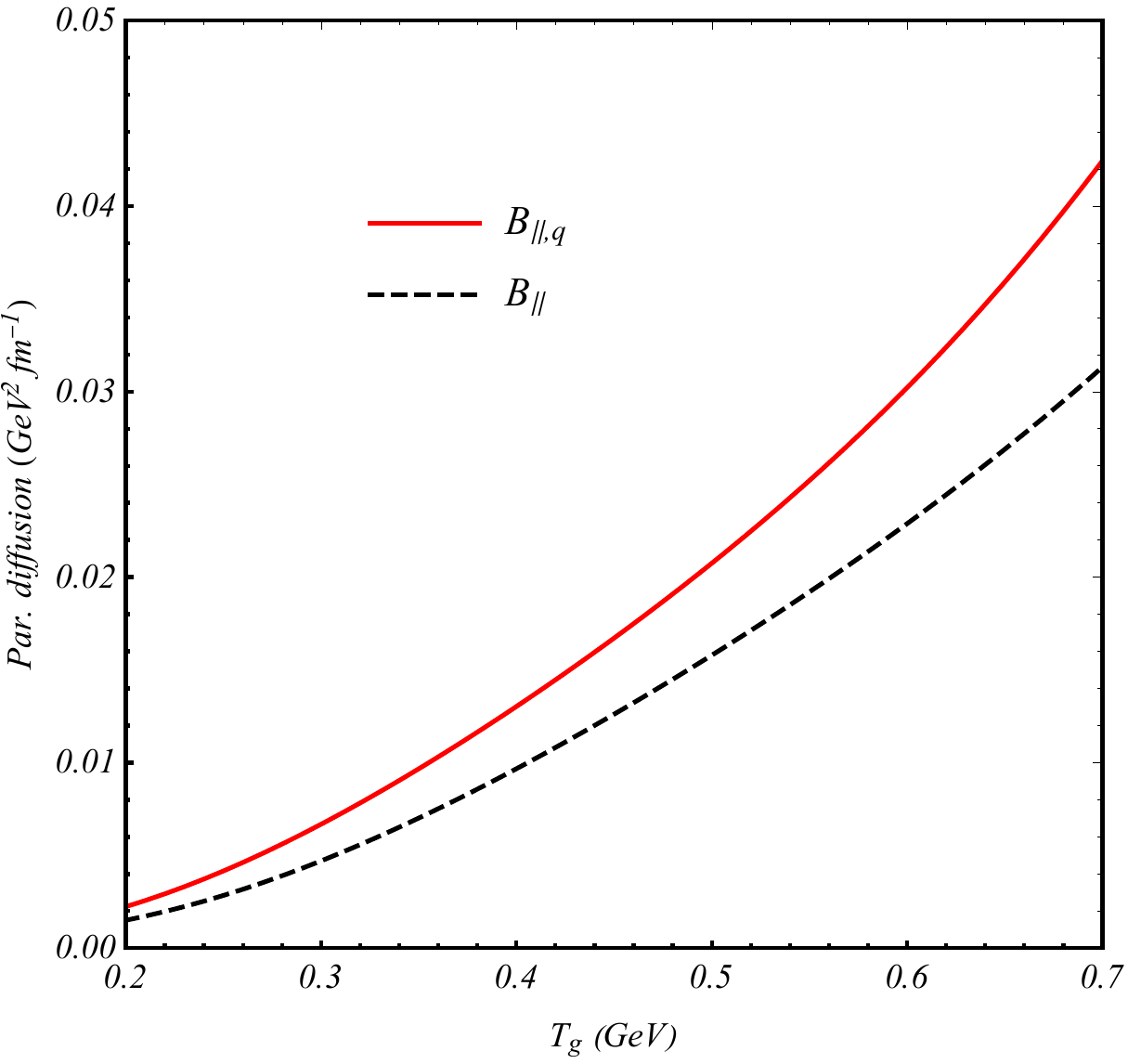}
\caption{Temperature variation of the nonextensive and extensive parallel diffusion coefficients of the light quarks.}\label{diffparT}
\endminipage\hfill
\end{figure}
\begin{figure}[!htb]
\minipage{0.42\textwidth}
\includegraphics[width=\linewidth]{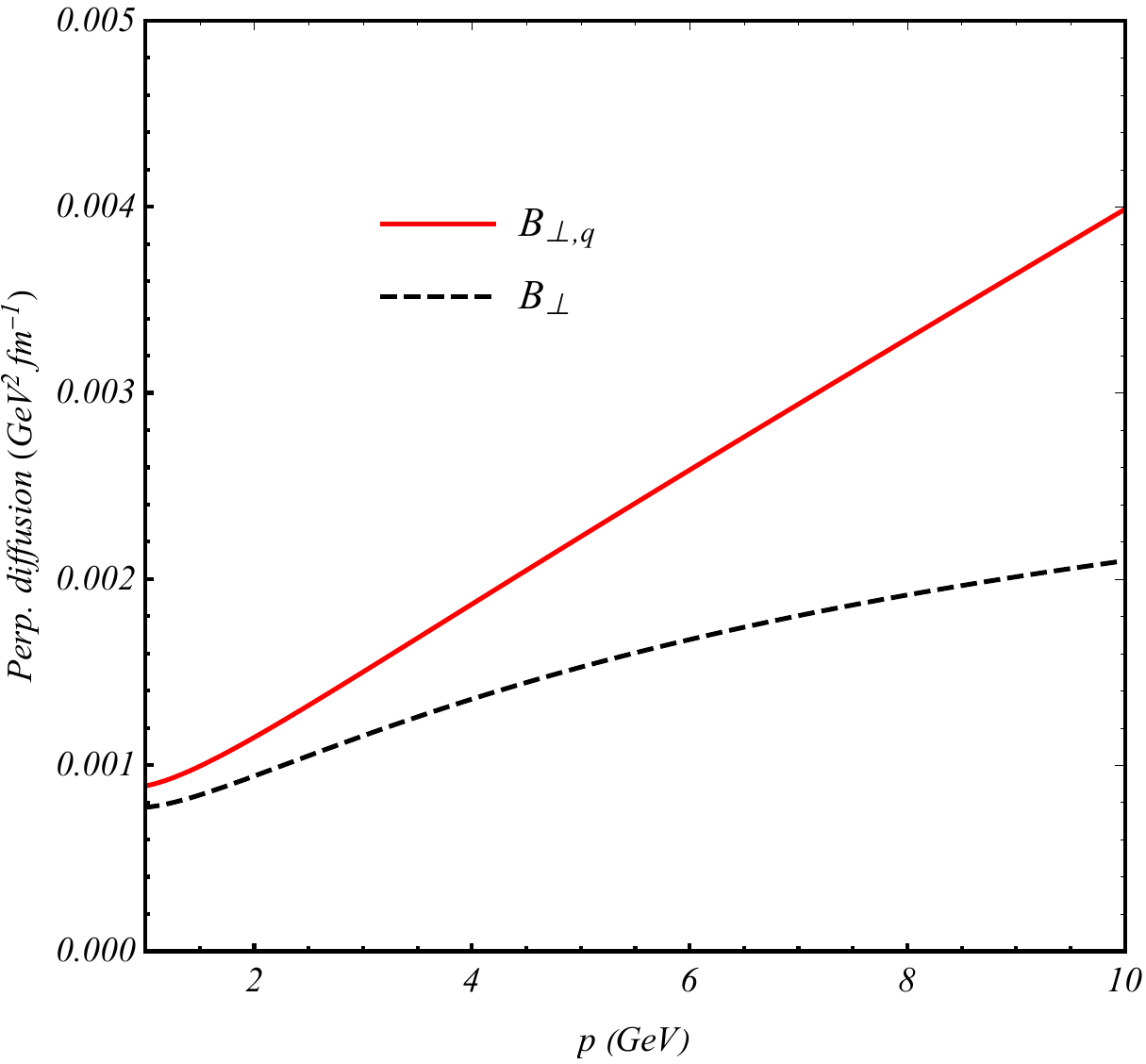}
\caption{Momentum variation of the nonextensive and extensive perpendicular diffusion coefficients of the light quarks.}\label{diffperpp}
\endminipage\hfill
\minipage{0.42\textwidth}
\hspace*{-1cm}
\includegraphics[width=\linewidth]{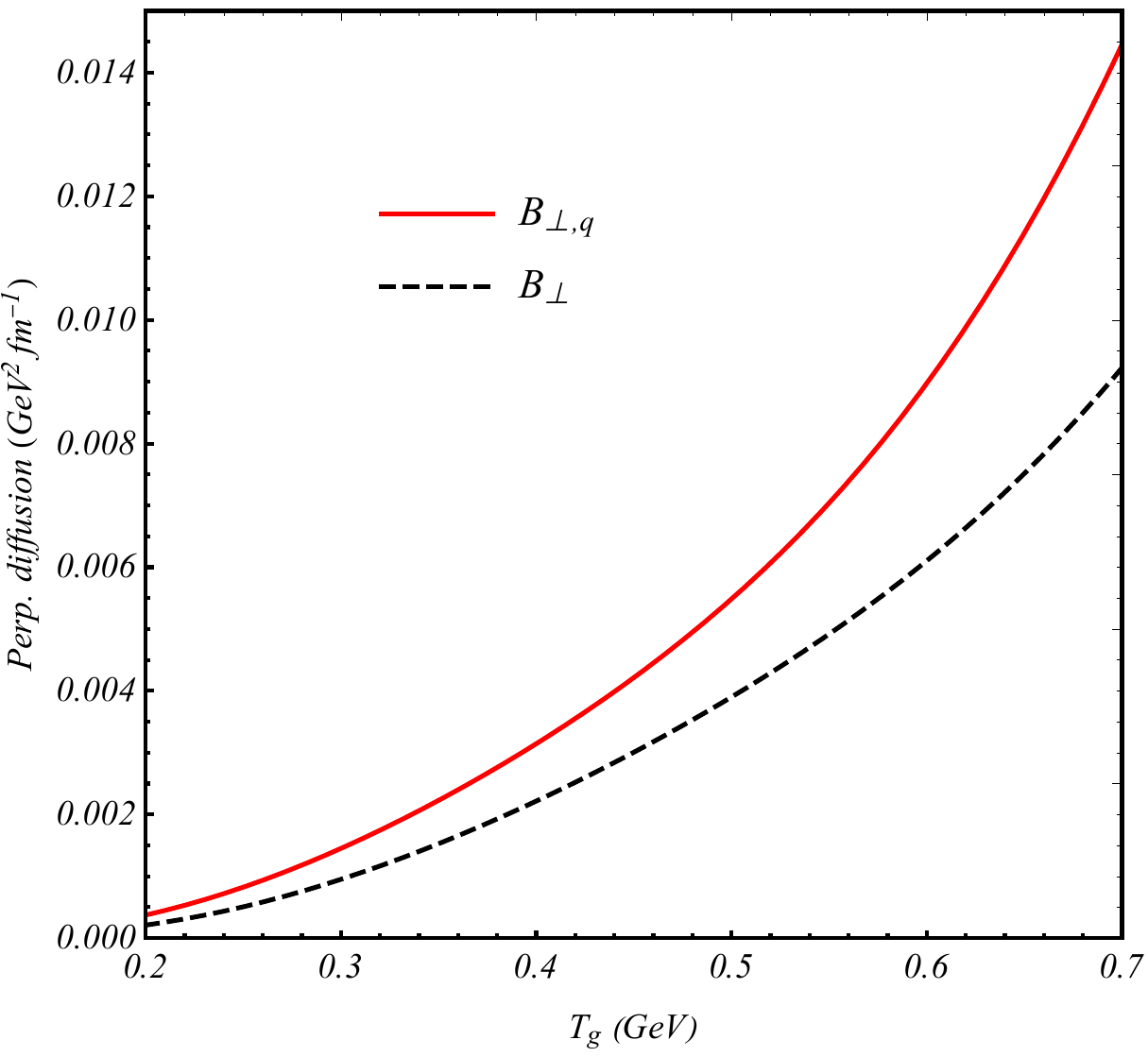}
\caption{Temperature variation of the nonextensive and extensive perpendicular diffusion coefficients of the light quarks.}\label{diffperpT}
\endminipage\hfill
\end{figure}

In this section we estimate the nonextensive Fokker-Planck drag and diffusion coefficients of energetic light quarks passing through a
gluonic plasma with the help of Eq.~\eqref{tslightdrdiff}. We express the nonextensive drag and diffusion coefficients in the following way,
\bea
A_{i,q} &=& \frac{P_i}{|\vec{P}|^2} A_q(|\vec{P}|^2), \nn\\
B_{ij,q} &=& \left(\delta_{ij}-\frac{{P_i}{P_j}}{{|\vec{P}|^2}}\right) B_{\bot,q} ({|\vec{P}|^2})+ 
\frac{{P_i}{P_j}}{{|\vec{P}|^2}} B_{||,q} (|\vec{P}|^2),
\eea
and evaluate $A_q$, $B_{\bot,q}$, and $B_{||,q} $ which, from Eq.~\eqref{tslightdrdiff}, can be evaluated to be,
\bea
A_q &=& \left<\left<1-\frac{\vec{P}.\vec{P'}}{|\vec{P}|^2}\right>\right>_q, \nn\\
B_{\bot,q} &=& \frac{f^{q-1}}{4} \left<\left<|\vec{P}|'^2 -\frac{(\vec{P}.\vec{P}')^2}{|\vec{P}|^2}\right>\right>_q, \nn\\
B_{||,q} &=& \frac{f^{q-1}}{2} \left<\left<\frac{(|\vec{P}|^2-\vec{P}.\vec{P}')^2}{|\vec{P}|^2}\right>\right>_q.
\eea
In the limit $q\rightarrow1$ they converge with the results given by the Boltzmann-Gibbs statistics. The momentum and 
temperature variation of the quantities are given in Figs. \ref{dragp}-
\ref{diffperpT}. We have taken $q=1.01$, and temperature to be 350 MeV (for Figs. \ref{dragp}, \ref{diffparp}, \ref{diffperpp}),
and momentum of the energetic light quark to be 5 GeV (for Figs. \ref{dragT}, \ref{diffparT}, \ref{diffperpT}).

\section{Summary, conclusions, and outlook}

In summary, we have studied the nonextensive Boltzmann transport equation in different approximations. First we have proposed approximate
analytical iterative closed form solutions for the NEBTE in the relaxation time approximation and have indicated how the results can be used to
describe the nuclear suppression factor (for a similar work using the conventional Boltzmann transport equation, see Ref. \cite{tsallisraaepja}).

We have observed that for a considerably wide range of the parameter values relevant for the high-energy physics phenomenology the first
order solution is very close to the exact solution. However, this approximated solution works better for higher mass particles. For the lower mass 
region ($\sim$140 MeV), slight deviation is observed at very low transverse momentum region ($\lesssim$0.6 GeV). The extent of agreement can 
be increased considering higher order solutions. 

From Figs. \ref{polar1}-\ref{chpiraa}, apart from the Tsallis parameter and temperature, we can get an estimate
of the ratio of the time scales relevant for evolution processes. We notice that the ratio of the freeze-out time to the relaxation time
($t/\tau$) has a value of the order of 1. An estimate of this quantity has recently been done in Ref. \cite{maciej} using the conventional Boltzmann transport equation which finds this value to be of the order of 1.5 assuming that the average transverse momentum remains constant during evolution.

We observe that though the Tsallis transport coefficients qualitatively follow the trends of the Boltzmann-Gibbs transport coefficients 
for the given momentum and temperature range, the former have higher numerical values. These higher values of transport coefficients 
can be attributed to the `interplay' between the test particle distribution and the medium distribution introduced by modifying the molecular chaos
hypothesis which is not valid for systems with relatively small number of particles \cite{deppmanqideal}. Given the indication for the QGP formation
in small systems \cite{natureALICE} where a generalized molecular chaos hypothesis may be important, the present calculations will be useful to 
characterize the hot and dense medium. Also, in this paper we have considered only the collisional processes. Inclusion of the radiative processes 
will be an interesting extension. One can also calculate the stopping power ($dE/dx$) from the present calculations by combining drag and $dE/dx$
in terms of a relativistically invariant quantity \cite{rafelskiwaltonprl}. 

\vspace{-5pt}
\section*{Acknowledgement}
The author acknowledges partial support from the joint project between the JINR and IFIN-HH.

\end{document}